\documentstyle[12pt]{article}
\newcommand{\nc}{\newcommand}
\nc{\postscript}[2] 
{\setlength{\epsfxsize}{#2\hsize}\centerline{\epsfbox{#1}}}
\nc{\bg}{B. Grz\c{a}dkowski}
\nc{\non}{\nonumber}
\nc{\barx}{\bar{x}}\nc{\pbarn}{\;\hbox {pb}}\nc{\fbarn}{\;\hbox {fb}}
\nc{\hc}{\hbox {h.c.}} \nc{\re}{\hbox {Re}} 
\nc{\mev}{\hbox {MeV}} \nc{\gev}{\;\hbox {GeV}}
\def\gesim{\lower0.5ex\hbox{$\:\buildrel >\over\sim\:$}} 
\def\lesim{\lower0.5ex\hbox{$\:\buildrel <\over\sim\:$}} 
\nc{\prd}[3]{{\it Phys.\ Rev.}\ {{\bf D{#1}} (#2), #3}}
\nc{\prl}[3]{{\it Phys.\ Rev.\ Lett.}\ {{\bf {#1}} (#2), #3}}
\nc{\plb}[3]{{\it Phys.\ Lett.}\ {{\bf B{#1}} (#2), #3}}
\nc{\npb}[3]{{\it Nucl.\ Phys.}\ {{\bf B{#1}} (#2), #3}}
\nc{\ptp}[3]{{\it Prog.\ Theor.\ Phys.}\ {{\bf {#1}} (#2), #3}}
\nc{\zfp}[3]{{\it Z.\ Phys.}\ {{\bf C{#1}} (#2), #3}}
\nc{\mpla}[3]{{\it Mod.\ Phys.\ Lett.}\ {{\bf A{#1}} (#2), #3}}
\nc{\rmp}[3]{{\it Rev.\ Mod.\ Phys.}\ {{\bf {#1}} (#2), #3}}
\nc{\ijmpa}[3]{{\it Int.\ J.\ of\ Mod.\ Phys.}\
               {{\bf A{#1}} (#2), #3}}
\nc{\ttbar}{t\bar{t}}         \nc{\bbbar}{b\bar{b}}
\nc{\tanb}{\tan \beta}        \nc{\twbdec}{t\to W^+ b}
\nc{\tbwbdec}{\bar{t}\to W^- \bar{b}}
\nc{\epem}{e^+e^-}            \nc{\eett}{\epem \to \ttbar}
\nc{\sigeett}{\sigma_{e\bar{e}\to\ttbar}}
\nc{\wpwm}{W^+W^-}            \nc{\tbar}{\bar{t}}
\nc{\bbar}{\bar{b}}           \nc{\wpp}{W^+}
\nc{\mt}{m_t}    \nc{\mts}{m_t^2}   \nc{\mw}{m_W}    \nc{\mws}{m_W^2}
\nc{\mz}{m_Z}    \nc{\mzs}{m_Z^2}
\nc{\ttbardec}{\ttbar \to W^+W^-\bbbar}
\nc{\wwbb}{W^+W^-\bbbar}      \nc{\sm}{SM}
\nc{\cw}{\cos\theta_W}        \nc{\sw}{\sin\theta_W}
\nc{\sws}{\sin^2\theta_W}     \nc{\sig}{\sigma_{tot}}
\nc{\lp}{\ell^+}              \nc{\lm}{\ell^-}
\nc{\epsl}{\epsilon_L}        \nc{\cp}{C\!P}

\nc{\splus}{s_+}       \nc{\smin}{s_-}        \nc{\eps}{\epsilon}
\nc{\psp}{Ps_+}        \nc{\psm}{Ps_-}        \nc{\lsp}{ls_+}
\nc{\lsm}{ls_-}        \nc{\sss}{s_+s_-}      \nc{\m}{m_t}
\nc{\mq}{m_t^2}        \nc{\mr}{\frac{1}{\m}} \nc{\av}{A_{\gamma}}
\nc{\bv}{B_{\gamma}}   \nc{\az}{A_Z}          \nc{\bz}{B_Z}
\nc{\avs}{A_{\gamma}^2}\nc{\azs}{A_Z^2}       \nc{\bzs}{B_Z^2}
\nc{\dav}{\delta \! A_{\gamma}}   \nc{\dbv}{\delta \! B_{\gamma}}
\nc{\dcv}{\delta C_{\gamma}}      \nc{\ddv}{\delta \! D_{\gamma}}
\nc{\daz}{\delta \! A_Z}          \nc{\dbz}{\delta \! B_Z}
\nc{\dcz}{\delta C_Z}             \nc{\ddz}{\delta \! D_Z}
\nc{\dev}{\delta \! E_{\gamma}}   \nc{\dez}{\delta \! E_Z}
\nc{\dfv}{\delta \! F_{\gamma}}   \nc{\dfz}{\delta \! F_Z}
\nc{\rdav}{{\rm Re}(\delta \! A_{\gamma}) \:}
\nc{\rdbv}{{\rm Re}(\delta \! B_{\gamma}) \:}
\nc{\rdcv}{{\rm Re}(\delta C_{\gamma}) \:}
\nc{\rddv}{{\rm Re}(\delta \! D_{\gamma}) \:}
\nc{\rdaz}{{\rm Re}(\delta \! A_Z) \:}
\nc{\rdbz}{{\rm Re}(\delta \! B_Z) \:}
\nc{\rdcz}{{\rm Re}(\delta C_Z) \:}
\nc{\rddz}{{\rm Re}(\delta \! D_Z) \:}
\nc{\idav}{{\rm Im}(\delta \! A_{\gamma}) \:}
\nc{\idbv}{{\rm Im}(\delta \! B_{\gamma}) \:}
\nc{\idcv}{{\rm Im}(\delta C_{\gamma}) \:}
\nc{\iddv}{{\rm Im}(\delta \! D_{\gamma}) \:}
\nc{\idaz}{{\rm Im}(\delta \! A_Z) \:}
\nc{\idbz}{{\rm Im}(\delta \! B_Z) \:}
\nc{\idcz}{{\rm Im}(\delta C_Z) \:}
\nc{\iddz}{{\rm Im}(\delta \! D_Z) \:}
\nc{\cz}{(1+v_e^2)d\:\!'^2}         \nc{\ci}{v_ed\:\!'}
\nc{\ccz}{v_ed\:\!'^2}              \nc{\cci}{d\:\!'}
\nc{\lspace}{\;\;\;\;\;\;\;\;\;\;}  \nc{\llspace}{\lspace \lspace}

\nc{\beq}{\begin{equation}}   \nc{\eeq}{\end{equation}}
\nc{\bea}{\begin{eqnarray}}   \nc{\eea}{\end{eqnarray}}
\nc{\baa}{\begin{array}}      \nc{\eaa}{\end{array}}
\nc{\bit}{\begin{itemize}}    \nc{\eit}{\end{itemize}}
\nc{\ben}{\begin{enumerate}}  \nc{\een}{\end{enumerate}}
\nc{\bce}{\begin{center}}     \nc{\ece}{\end{center}}
\begin{document}
\pagestyle{empty} \setlength{\footskip}{2.0cm}
\setlength{\oddsidemargin}{0.5cm} \setlength{\evensidemargin}{0.5cm}
\renewcommand{\thepage}{-- \arabic{page} --}
\def\mib#1{\mbox{\boldmath $#1$}}
\def\bra#1{\langle #1 |}      \def\ket#1{|#1\rangle}
\def\vev#1{\langle #1\rangle} \def\dps{\displaystyle}
% -------------------------------------------------------------------
   \def\thebibliography#1{\centerline{REFERENCES}
     \list{[\arabic{enumi}]}{\settowidth\labelwidth{[#1]}\leftmargin
     \labelwidth\advance\leftmargin\labelsep\usecounter{enumi}}
     \def\newblock{\hskip .11em plus .33em minus -.07em}\sloppy
     \clubpenalty4000\widowpenalty4000\sfcode`\.=1000\relax}\let
     \endthebibliography=\endlist
   \def\sec#1{\addtocounter{section}{1}\section*{\hspace*{-0.72cm}
     \normalsize\bf\arabic{section}.$\;$#1}\vspace*{-0.3cm}}
% -------------------------------------------------------------------
\vspace*{-1.3cm}\noindent
\hspace*{11.cm}TOKUSHIMA 99-02\\
\hspace*{11.cm}(hep-ph/9908345)\\

\vspace*{1.3cm}

\begin{center}
\renewcommand{\thefootnote}{*}
{\large\bf Probing Anomalous Top-Couplings at Polarized}

\vskip 0.15cm
{\large\bf Linear Collider}$\,$\footnote{Talk at the fourth
International Workshops on Linear Colliders (LCWS99), April 28 - May
5, 1999, Sitges, Barcelona, Spain. \\
This work is based on collaboration with \bg.}
\end{center}

\vspace*{1.2cm}
\renewcommand{\thefootnote}{*)}
\begin{center}
{\sc Zenr\=o HIOKI$^{\:}$}\footnote{E-mail address:
\tt hioki@ias.tokushima-u.ac.jp}
\end{center}

\vspace*{1.2cm}
\centerline{\sl Institute of Theoretical Physics,\ 
University of Tokushima}

\vskip 0.14cm
\centerline{\sl Tokushima 770-8502, JAPAN}

\vspace*{3cm}
\centerline{ABSTRACT}

\vspace*{0.4cm}
\baselineskip=20pt plus 0.1pt minus 0.1pt
The energy spectra of the lepton(s) in $e\bar{e}\to t\bar{t}\to
\ell^\pm X/\ell^+\ell^- X'$ at next linear colliders (NLC) are
analyzed a model-independent way for arbitrary longitudinal beam
polarizations as a general test of possible anomalous top-quark
couplings.

\vfill
\newpage
%--------------------------------------------------------------------
\renewcommand{\thefootnote}{\sharp\arabic{footnote}}
%--------------------------------------------------------------------
\pagestyle{plain} \setcounter{footnote}{0}
\baselineskip=21.0pt plus 0.2pt minus 0.1pt

% 1111111111111111111111111111111111111111111111111111111111111111111
\sec{Introduction}

A lot of data have been accumulated on the top-quark since its
discovery. However it is still an open question whether its
interactions obey the standard scheme like all the other fermions or
there exists some new-physics contribution to its couplings. The top
quark decays immediately after being produced because of its huge
mass. Therefore the decay process is not influenced by any
hadronization effects and consequently its decay products are
expected to carry valuable information on the top properties. 

Next linear colliders (NLC) of $e\bar{e}$ will give us fruitful data
on the top through $e\bar{e}\to t\bar{t}$. In particular the energy
spectra of the lepton(s) produced in its semileptonic decay(s) turn
out to be a useful analyzer of the top-quark couplings \cite{SP}.
Indeed many authors have worked on this subject (see the reference
list of Ref.[2]), and we also have tackled them over the past several
years. Here I would like to show some of the results of our latest
model-independent analyses \cite{GH} via arbitrary longitudinal beam
polarizations, where we have assumed the most general anomalous
couplings both in the production and decay vertices in contrast to
most of the existing works.
  
% 2222222222222222222222222222222222222222222222222222222222222222222
\sec{Framework}

We can represent the most general $\ttbar$ couplings to the photon
and $Z$ boson as
\begin{equation}
{\mit\Gamma}_{vt\bar{t}}^{\mu}=
\frac{g}{2}\,\bar{u}(p_t)\,\Bigl[\,\gamma^\mu \{A_v+\delta\!A_v
-(B_v+\delta\!B_v) \gamma_5 \}
+\frac{(p_t-p_{\bar{t}})^\mu}{2m_t}(\delta C_v-\delta\!D_v\gamma_5)
\,\Bigr]\,v(p_{\bar{t}})  \label{ff}
\end{equation}
in the $m_e=0$ limit, where $g$ denotes the $SU(2)$ gauge coupling
constant, $v=\gamma,Z$, and
\[
\av=\frac{4}{3}\sw,\ \ \bv=0,\ \ \az=\frac{v_t}{2\cw},\ \ \bz
=\frac{1}{2\cw}
\]
with $v_t\equiv 1-(8/3)\sin^2\theta_W$. Among the above form factors,
$\delta\!A_{\gamma,Z}$, $\delta\!B_{\gamma,Z}$, $\delta C_{\gamma,Z}$
and $\delta\!D_{\gamma,Z}$ are parameterizing $C\!P$-conserving and
$C\!P$-violating non-standard interactions, respectively. 

On the other hand, we adopted the following parameterization of
the $Wtb$ vertex suitable for the $\twbdec$ and $\tbwbdec$ decays:
\begin{eqnarray}
&&\!\!{\mit\Gamma}^{\mu}_{Wtb}=-{g\over\sqrt{2}}\:
\bar{u}(p_b)\biggl[\,\gamma^{\mu}(f_1^L P_L +f_1^R P_R)
-{{i\sigma^{\mu\nu}k_{\nu}}\over M_W}
(f_2^L P_L +f_2^R P_R)\,\biggr]u(p_t),\ \ \ \ \ \ \label{ffdef}\\
&&\!\!\bar{\mit\Gamma}^{\mu}_{Wtb}=-{g\over\sqrt{2}}\:
\bar{v}(p_{\bar{t}})
\biggl[\,\gamma^{\mu}(\bar{f}_1^L P_L +\bar{f}_1^R P_R)
-{{i\sigma^{\mu\nu}k_{\nu}}\over M_W}
(\bar{f}_2^L P_L +\bar{f}_2^R P_R)\,\biggr]v(p_{\bar{b}}),
\label{ffbdef}
\end{eqnarray}
where $k$ is the momentum of $W$ and $P_{L/R}=(1\mp\gamma_5)/2$. It
is worth to mention that the form factors for top and anti-top
satisfy the following relations \cite{cprelation}:
\begin{equation}
f_1^{L,R}=\pm\bar{f}_1^{L,R},\lspace f_2^{L,R}=\pm\bar{f}_2^{R,L},
\label{cprel}
\end{equation}
where upper (lower) signs are those for $C\!P$-conserving
(-violating) contributions.

For the initial beam-polarization we used the following convention:
\begin{eqnarray}
&&P_{e^-}=+[N(e^-,+1)-N(e^-,-1)]/[N(e^-,+1)+N(e^-,-1)],\\
&&P_{e^+}=-[N(e^+,+1)-N(e^+,-1)]/[N(e^+,+1)+N(e^+,-1)],
\end{eqnarray}
where $N(e^{-(+)},h)$ is the number of $e^-(e^+)$ with helicity $h$
in each beam.

% 3333333333333333333333333333333333333333333333333333333333333333333
\sec{Lepton-energy spectra}

After some calculations, we arrived at the normalized single
distribution, which we express as
\beq
\frac1{\sigma}{\frac{d\sigma}{dx}}=\sum_{i=1}^{3}c_i^\pm f_i(x).
\label{single}
\eeq
Here $\pm$ corresponds to $\ell^{\pm}$ and the variable $x$ is
defined from the top velocity $\beta$ and the lepton energy\ 
$E_{\ell}$, both in the $e\bar{e}$ c.m. frame, as \cite{AS}
$$
x\equiv
\frac{2 E_\ell}{\mt}\left(\frac{1-\beta}{1+\beta}\right)^{1/2}.
$$
The coefficients $c_i^\pm$ on the right-hand side are given by
$$
c_1^\pm =1,
$$
$$
c_2^\pm = a_1\,\delta\!D_V^{(*)}
-a_2\,[\,\delta\!D_A^{(*)}-{\rm Re}(G_1^{(*)})\,]+a_3{\rm Re}
(\delta\!D_{V\!\!A}^{(*)})\mp \xi^{(*)},
$$
$$
c_3^+={\rm Re}(f_2^R),\ \ c_3^-={\rm Re}(\bar{f}_2^L),
$$
where $\delta\!D_{V,A,V\!\!A}^{(*)}$ and $G_1^{(*)}$ in $c_2^\pm$ are
combinations of the SM and non-SM form factors in eq.(\ref{ff}) but
without $\delta\!D_v$ while $\xi^{(*)}$ is one including $\delta\!
D_v$. This means $\xi^{(*)}$ is a parameter to express $C\!P$\ 
violation in the $t\bar{t}v$ couplings, and that is why the signs of
the $\xi^{(*)}$ terms for $\ell^+$ and $\ell^-$ are opposite to each
other. On the other hand, the coefficients $a_{1,2,3}$ consist of the
SM parameters only, and $f_{1,2,3}(x)$ are analytic functions
calculated in the SM.

Similarly, the normalized double lepton-energy spectrum is given by
the following formula:
\beq
\frac1{\sigma} 
\frac{d^2 \sigma}{dx d\bar{x}}\;=\;\sum_{i=1}^6 c_i f_i(x,\bar{x}),
\label{double}
\eeq
where $x$ and $\bar{x}$ are for $\ell^+$ and $\ell^-$ respectively,
\[
c_1=1,\;\;c_2=\xi^{(*)},\;\;c_3=\frac12{\rm Re}(f_2^R-\bar{f}_2^L),
\]
\[
c_4=a_1'\,\delta\!D_V^{(*)}+a_2'\,\delta\!D_A^{(*)}
+a_3'{\rm Re}(G_1^{(*)}),
\]
\[
c_5 = a_1\,\delta\!D_V^{(*)}
-a_2\,[\,\delta\!D_A^{(*)}-{\rm Re}(G_1^{(*)})\,]+a_3{\rm Re}
(\delta\!D_{V\!\!A}^{(*)}),
\]
\[
c_6=\frac12{\rm Re}(f_2^R+\bar{f}_2^L),
\]
and again $a_{1,2,3}'$ are combinations of the SM parameters and
$f_i(x,\bar{x})$ are analytic functions derived in the SM.

% 4444444444444444444444444444444444444444444444444444444444444444444
\sec{Parameter determination}

In order to study how precisely we can determine the coefficients $
c_i$ in eqs.(\ref{single},\ref{double}) when we have $N$\ 
corresponding events, we used the optimal observable procedure 
\cite{optimalization}. According to its prescription, we can deduce $
c_i$ from the spectra (which we express as ${\mit\Sigma(\phi)}$, 
where $\phi$ means $x$ or $(x, \bar{x})$) with statistical
uncertainty
$$
{\mit\Delta}c_i=\sqrt{X_{ii}/N},
$$
where $X_{ij}$ is the inverse matrix of
$$
M_{ij}=\int d\phi\frac{f_i(\phi)f_j(\phi)}{{\mit\Sigma(\phi)}}.
$$

From the theoretical point of view, perfectly-polarized beams ($
P_{e^+}=P_{e^-}=\pm 1$) are the most attractive. However, those are
difficult to achieve in practice, especially for the positron beam.
So we discussed the following two cases:

\hskip 2cm
(1)\ \ 
$P_{e^+}=0\,$\ \ vs\ \ $P_{e^-}=0,\ \pm 0.5,\ \pm 0.8$\ \ and\ \ $\pm
1$,

\hskip 2cm
(2)\ \ 
$P_{e^+}=P_{e^-}(\equiv P_e)=0,\ \pm 0.5,\ \pm 0.8$\ \ and\ \ $\pm
1$.

\noindent
In the analyses we assumed $\epsilon_{\ell}=0.6$ as the
lepton-tagging efficiency and $L=100$~fb$^{-1}$ as the integrated
luminosity.

Now let me show the main feature of the results focusing on the
single spectrum: we found both ${\mit\Delta}c^\pm_{2,3}$ become
smallest for $P_{e^-}=-1$/$P_e=-1$. We can thereby conclude
immediately that the best precision is obtained at $c_3$ measurements
for these polarizations. However we have to be a bit more careful for
$c_2$ measurements. This is because $c_2$ themselves vary depending
on the polarization. Therefore we should discuss the statistical
significance $N_{S\!D}\equiv |c_2^\pm|/{\mit\Delta}c^\pm_2$
inevitably instead of statistical errors only. For this purpose we
considered the following two sets of the couplings as an example:
\begin{itemize}
\item[(a)] 
 ${\rm Re}(\delta\!A_{\gamma,Z})={\rm Re}(\delta\!B_{\gamma,Z})
 ={\rm Re}(\delta  C_{\gamma,Z})={\rm Re}(\delta\!D_{\gamma,Z})=0.1$,
\item[(b)] 
 ${\rm Re}(\delta\!A_\gamma)={\rm Re}(\delta\!B_\gamma)
 ={\rm Re}(\delta  C_\gamma)={\rm Re}(\delta\!D_\gamma)=0.1$,\\
 ${\rm Re}(\delta\!A_Z)={\rm Re}(\delta\!B_Z)
 ={\rm Re}(\delta  C_Z)={\rm Re}(\delta\!D_Z)=-0.1$.
\end{itemize}

As a result, we found that the use of negatively-polarized beam(s) is
not always optimal: for the parameter set (a) a good precision in $
c_2^+$ measurements is obtained when $P_e<0$, but even in this case
the precision in $c_2^-$ measurements becomes better for $P_e>0$ or
even $P_e=0$. Moreover in case (b) both $c_2^+$ and $c_2^-$ get the
highest precision for $P_e=+1$. Therefore one should carefully adjust
optimal polarization to test any given model.

In any case one can conclude that (as far as the coefficient sets
discussed here are concerned) appropriate beam polarization(s)
provides measurements of $c_{2,3}^\pm$ at least at $2\sigma$ and $
3\sigma$ level for $P_{e^+}=0$ and $P_{e^+}\neq 0$, respectively
except for $c_2^-$ in case (a), where $|c_2^-|$ becomes tiny due to
an accidental cancellation.

We reached a similar conclusion for the double spectrum too: It
depends on the structure of tested models what polarization(s) is
best to study $t\bar{t}v$-couplings. I skip showing the details here,
however, for want of space.

% 5555555555555555555555555555555555555555555555555555555555555555555
\sec{Summary}

Next-generation linear colliders of $e^+ e^-$, NLC, are expected to
work as the cleanest facilities for studying top-quark interactions.
There, we will be able to perform detailed tests of the top-quark
couplings to the vector bosons and either confirm the SM simple
generation-repetition pattern or discover some non-standard
interactions. In this talk, I have shown main points of our latest
model-independent analyses of the single- and the double-leptonic
energy spectra for arbitrary longitudinal beam 
polarizations \cite{GH}.

We found ($i$) the use of longitudinal beams could be very effective
in order to increase precision of the determination of non-SM
couplings. However ($ii$) optimal polarization depends on the model
of new physics under consideration. Therefore polarization of the
initial beams should be carefully adjusted for each tested model.

\vspace*{0.6cm}
% RRRRRRRRRRRRRRRRRRRRRRRRRRRRRRRRRRRRRRRRRRRRRRRRRRRRRRRRRRRRRRRRRRR


\begin{thebibliography}{99}
%
\bibitem{SP} C.R. Schmidt and M.E. Peskin, \prl{69}{1992}{410}.

\bibitem{GH} \bg\ and Z. Hioki, Report IFT-06-98 -- TO\-KU\-SHI\-MA
98-02 (hep-ph/9805318).

\bibitem{AS} T. Arens and L.M. Sehgal, \prd{50}{1994}{4372}.

\bibitem{cprelation} W. Bernreuther, O. Nachtmann, P. Overmann and 
T. Schr\"{o}der, \npb{388}{1992}{53};\\
\bg\ and J.F. Gunion, \plb{287}{1992}{237}.

\bibitem{optimalization} 
J.F. Gunion, \bg\ and X-G. He, \prl{77}{1996}{5172}.
%
\end{thebibliography}
\end{document}